\documentclass[useAMS,usenatbib,referee]{mn2e}
\usepackage{graphicx}
\title[Observing Site Ranking]{Astronomical Site Ranking Based on Tropospheric
Wind Statistics}

\author[Garc\'{\i}a-Lorenzo et al.]{B. Garc\'{\i}a-Lorenzo$^{1}$\thanks{E-mail:
bgarcia@iac.es}, J.J. Fuensalida$^{1}$, C. Mu\~noz-Tu\~n\'on$^{1}$, and E. Mendizabal$^{1,2}$ \\
1 Instituto de Astrof\'{\i}sica de Canarias, C/Via Lactea S/N, 38350-La Laguna, Tenerife, Spain\\
2 Instituto Nacional de Astrof\'{\i}sica, \'Optica y Electr\'onica, C/Luis Enrique Erro,1, Tonantzintla, Puebla 72840, Mexico\\}

\begin{document}

\date{Accepted ..... Received .....; in original form .....}

\pagerange{\pageref{firstpage}--\pageref{lastpage}} \pubyear{2004}

\maketitle

\label{firstpage}

\begin{abstract} 

High altitude wind speeds  have been adopted as a parameter for astronomical 
site 
selection based on the relationship found at the Paranal and Cerro Pach\'on sites 
between the average velocity of the turbulence ($V_{0}$) and winds at the 
200 millibar pressure level ($V_{200}$). Although this relationship has not 
been checked at any other site in the world and a connection between image quality 
and $V_{200}$ has not been proven anywhere, high altitude wind speed ($V_{200}$) is a 
parameter for checking the suitability of sites for adaptive optics and surveying 
potential sites for ELT.

We present  comprehensive and reliable statistics of high altitude wind speeds 
and the tropospheric flows at the location of five important astronomical 
observatories. We have used the National Center for Environmental Prediction 
(NCEP/NCAR) Reanalysis database to collect wind data at nine tropospheric pressure 
levels for the five selected sites. For comparison and validation of the 
data from the climate diagnostic model, we have also obtained wind profiles from 
radiosonde stations. The degrees of correlation found indicate a high level of 
significance between NCEP/NCAR Reanalysis and balloon databases, pointing to 
NCEP/NCAR Reanalysis as a useful database for site characterization. 

 Statistical analysis exclusively of high altitude winds point to La Palma as 
 the most 
 suitable site for adaptive optics, 
 with a mean value of 22.13 m s$^{-1}$ at the 200 mbar pressure level. La Silla is at 
 the bottom of the ranking, with the largest average value 200 mbar wind speed
 (33.35 m s$^{-1}$). We have 
 found a clear annual periodicity of high altitude winds for the five sites in 
 study.  

We have also explored the connection of high to low altitude atmospheric winds as a 
first approach of the linear relationship between the average velocity of the 
turbulence and high altitude winds (Sarazin \& Tokovinin 2001). We may conclude 
that high and low altitude winds show good linear relationships at the five 
selected sites. The highest correlation coefficients correspond to Paranal and 
San Pedro M\'artir, while La Palma and La Silla show similar high to low altitude 
wind connection. Mauna Kea shows the smallest degree of correlation, which suggests a 
weaker linear relationship. Our results support the idea of  high 
altitude winds as a parameter for rank astronomical sites in terms of
 their suitability 
for adaptive optics, although we have no evidence for adopting the same linear 
coefficient at different sites. The final value of this linear coefficient at a 
particular site could drastically change  the interpretation of high altitude 
wind speeds as a direct parameter for site characterization. 
\end{abstract}


\section{Introduction}

Ground-based stronomical observations are drastically affected by the presence of
atmospheric turbulence.  The theoretical angular resolution of a telescope, 
determined by
its diameter, is usually severely degraded because of atmospheric turbulence.
Adaptive optics techniques have been developed to compensate for the effects of the
atmosphere on  astronomical images and to reach the diffraction limit of our
telescopes.  However, the design and operation of adaptive optic systems are
complex and difficult because the turbulence above astronomical
sites is still insufficiently well characterized.  
In spite of partial information on the turbulence
behaviour above existing astronomical observatories, it is obvious that a site
with low and stable turbulence should be better suited for the implementation of
adaptive optics
 than  one with a turbulent and chaotic atmosphere.  The requirements for excellent
image quality  of current large and future very large telescopes demand
a proper knowledge of  atmospheric turbulence, and several projects are already
pursuing this aim.  Moreover, the precise characterization of the turbulence above
a particular site requires long term monitoring.  To counter the lack of
long term information on turbulence,  high altitude winds (in particular
winds at the 200 mbar pressure level---hereafter $V_{200}$) have been proposed
(Sarazin \&
Tokovinin 2001) as a
parameter for estimating the total turbulence at a particular site, 
 since records of this parameter exist from several sources.  This
proposal is based on the idea that the greatest source for turbulence generation is
related to the highest peak in the vertical wind profile, which is located at
the 200
mbar pressure level globally.  Also, Sarazin \& Tokovinin (2001) find a
good correlation between the average velocity of the turbulence, $V_{0}$, and
$V_{200}$ of the form $V_{0}=0.4V_{200}$ at the Cerro Pach\'on and Paranal
Observatories in Chile.  Although the $V_{0} \propto$ $V_{200}$ relationship has
not been tested at other sites, its validity would simplify the calculation of key
parameters for adaptive optics, such as the coherence time.  Furthermore, 
  $V_{200}$ statistics could be used as a parameter for ranking the
suitability  of different observing sites for AO. However, such a relationship 
never indicates a connection between the total turbulence at the ground level 
(seeing) and  wind speed at the 200 mbar pressure level. Indeed, Sarazin \& 
Tokovinin (2001) have also studied this option but a connection has not been
found. Despite 
 this, a false idea of a connection between image quality and high altitude 
wind speed is increasingly  widespread among those in the astronomical community 
interested in adaptive 
optics.

  Some studies centred on $V_{200}$
 statistics values have been recently published (Carrasco \& Sarazin 2003;
Sarazin 2002), and a detailed statistical analysis of $V_{200}$ for  Roque de
los Muchachos Observatory (ORM) on the island of La Palma (Spain) have been
performed by Chueca et al.\ (2004). While the linear $V_{0} \propto V_{200}$  
relationship (Sarazin \&
Tokovinin 2001) is confirmed at other sites, a first approach to connecting high
altitude winds to turbulence at ground level can be carried out by studying the wind
vertical profile and the relation of high to low altitude winds.  Previous studies
have also reported a connection of ground layer winds to image quality
(\citet{mu98}; \citet{var01}).

In this paper we present  a detailed statistical analysis of $V_{200}$ for
different observing sites in the two hemispheres.  For the first time, we study
the connection of high to low altitude winds and we present the statistical
analysis of the wind vertical profile at five astronomical sites, the ORM, Mauna Kea 
and San Pedro Martir in the north, and Paranal and La Silla in the south (Table
\ref{locations}).  We also discuss the suitability for adaptive optics of the five 
 summits analysed on the basis of the wind vertical profile statistics.

\section{ THE DATA}

We have selected five existing sites spread all over the globe to study the
behaviour of $V_{200}$ and the  vertical wind profile.  Table
\ref{locations} lists the astronomical observatories considered, their
location and altitude.  We used the National Center for Environmental Prediction
(NCEP/NCAR) Reanalysis database to collect wind speeds at different pressure
levels for the five sites.  The Reanalysis data were obtained as 6-hourly and
daily {\it U}-wind and {\it V}-wind components from the beginning of
 1980 to the end of 2002.
Wind speeds in this database are considered as one of the most reliably analysed fields
(see \cite{kal96} and \cite{kis01} for a detailed description) and they are heavily
constrained by observational data. 
eW
obtained the 6-hourly and daily wind speed module from {\it U}-wind and {\it V}-wind
 6-hourly
and daily components (more than 33\,600 data for each component and location).

\begin{table*}
\centering
 \begin{minipage}{180mm}
  \caption{Location of selected observing sites (from http://www.seds.org/billa/bigeyes.html)}
\begin{tabular}{l|ccccc}
                &  {\bf ORM(La Palma)} & {\bf La Silla}&{\bf Mauna Kea} & {\bf Paranal} &  {\bf San Pedro M\'artir} \\ \hline
{\bf Latitude}  &     28 46 N    &     29 15 S    &    19 50 N    &    24 38 S     &  31 02 N\\
{\bf Longitude} &     17 53 W    &     70 44 W    &    155 28 W   &    70 24 W     &  115 27 W\\
{\bf Altitude}  &     2400 m     &      2400 m    &    4100 m     &    2635 m      &  2800 m \\
{\bf Country}    &    Canary Islands, Spain & Chile &  Hawaii, USA & Chile & Baja California, Mexico\\ \hline
\end{tabular}\label{locations}
\end{minipage}
\end{table*}

For comparison  and validation of the data from the climate diagnostic
model, we have obtained winds from radiosonde stations.  The main criteria in
selecting radiosonde stations were the year range for available data and the
distance from the observing sites. Table \ref{stations} shows information about
the final selected stations.

\begin{table*}
\centering
 \begin{minipage}{180mm}
  \caption{Radiosonde stations closest to the selected astronomical sites.}
\begin{tabular}{l|cccccc}
{\bf Station} & {\bf Place} & {\bf Latitude} & {\bf Longitude} & {\bf Available} & {\bf Close to} & {\bf Distance to}\\ 
              &             &                &                 &   {\bf Range}   & {\bf Site} &    {\bf Site (km)}   \\\hline

600200 & Santa Cruz de Tenerife (Spain) & 28.467 N & 16.25 W & 1960-2001 & La Palma & 163 \\
855430 & Quintero (Chile)              & 32.780 S &  71.517 W & 1957-1999 & La Silla & 400 \\ 
912850 & Hilo (Hawaii, USA) & 19.716 N &  155.06 W & 1950-2002 & Mauna Kea & 44.5 \\
854420 & Antofagasta (Chile)           & 23.410 S &  70.467 W & 1957-2001 & Paranal & 136 \\
722930 & California (USA)              & 32.867 N &  117.15 W & 1972-2001 & San Pedro Martir & 260 \\ \hline
\end{tabular}\label{stations}
\end{minipage}
\end{table*}

\begin{figure}
\centering
\includegraphics[scale=0.6]{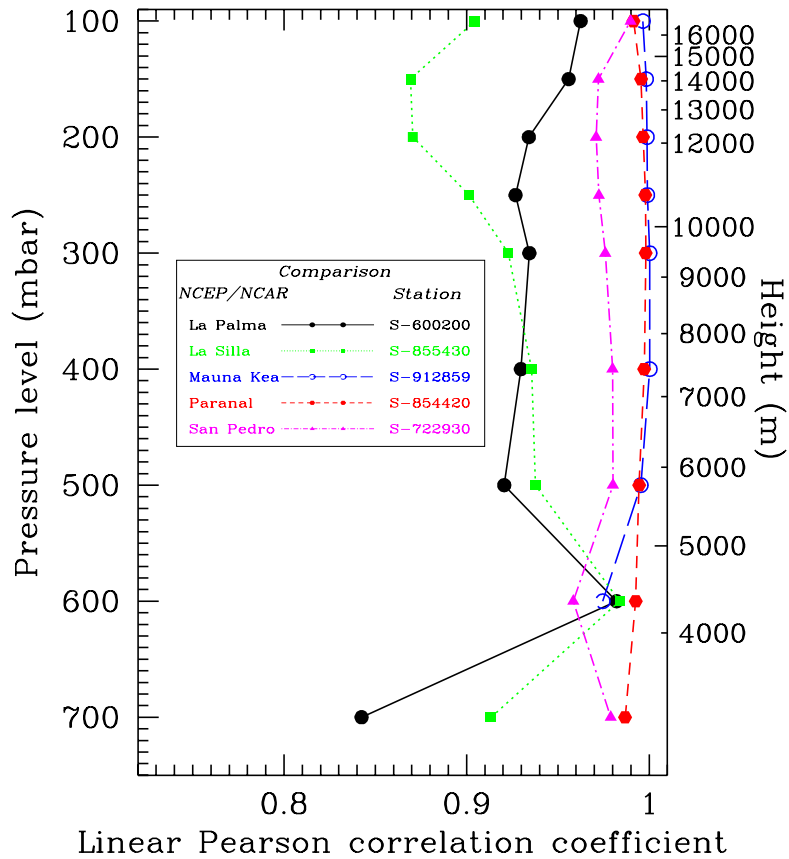}
 \caption{ Linear Pearson correlation coefficient obtained comparing the mean
  monthly wind statistics for the NCEP/NCAR data and balloon measurements at
   different pressure levels for the period 1980--2002. The height axes have
    been derived from the pressure to height relationship assuming ICAO standard
     atmosphere.}
\label{pearson}
\end{figure}

In both databases we have selected the simultaneous data for the period 1980--2002
to derive the mean monthly wind at the different common pressure levels.  We found
a good correlation between the balloon and NCEP/NCAR data statistic in the sites
 studied.  Figure \ref{pearson} shows the linear Pearson correlation coefficient
of the mean monthly $V_{200}$ time series from the two databases at different
pressure levels.  The degrees of correlation  found indicate a high level of
significance between the NCEP/NCAR Reanalysis and balloon databases at the five
selected locations and for any altitude considered, thereby revealing NCEP/NCAR
Reanalysis as a useful database for site characterization approach.  The detailed
analysis of winds in the following sections will be performed only with the
NCEP/NCAR Reanalysis data at each site.  This database provide a better temporal
coverage and resolution than the balloon measurements and have no gaps in the time
series

\section{Analysis}

Sarazin \& Tokovinin (2001) have suggested that the suitability of a site for
adaptive optics is related to winds at the pressure level of 200 mbar.  The
average altitude of this pressure level at the Earth's poles is about one
kilometre lower than at the equator.  Moreover, the 200 mbar pressure level
presents a slightly seasonal altitude dependency,  being about 400
m lower in altitude in winter than   in summer time.  All
 other pressure levels show a similar behaviour.  Assuming an ICAO standard
atmosphere (see  http://www.pdas.com/coesa.htm for an exact definition), we
shall consider hereafter a mean altitude of 12170 meters for the 200 mbar pressure
level at any site.  In the same way, we adopt the same mean altitude for any of
the pressure levels considerer in this paper at any site. Therefore, the height
axis of the figures can only be taken as an approximate value.

\subsection{$V_{200}$  statistics and seasonal behaviour}

\begin{figure}
\includegraphics[scale=0.4]{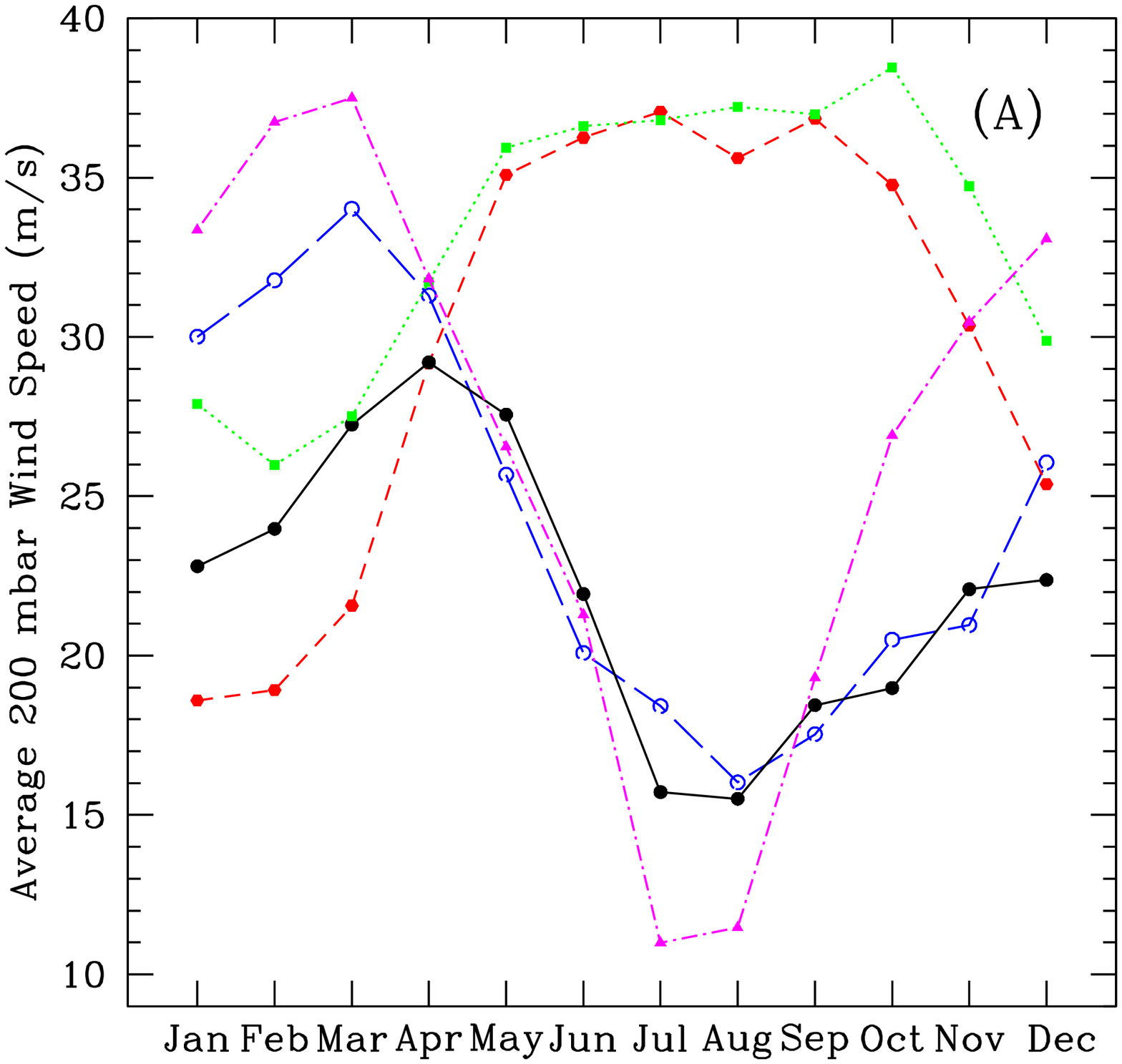}
\includegraphics[scale=0.4]{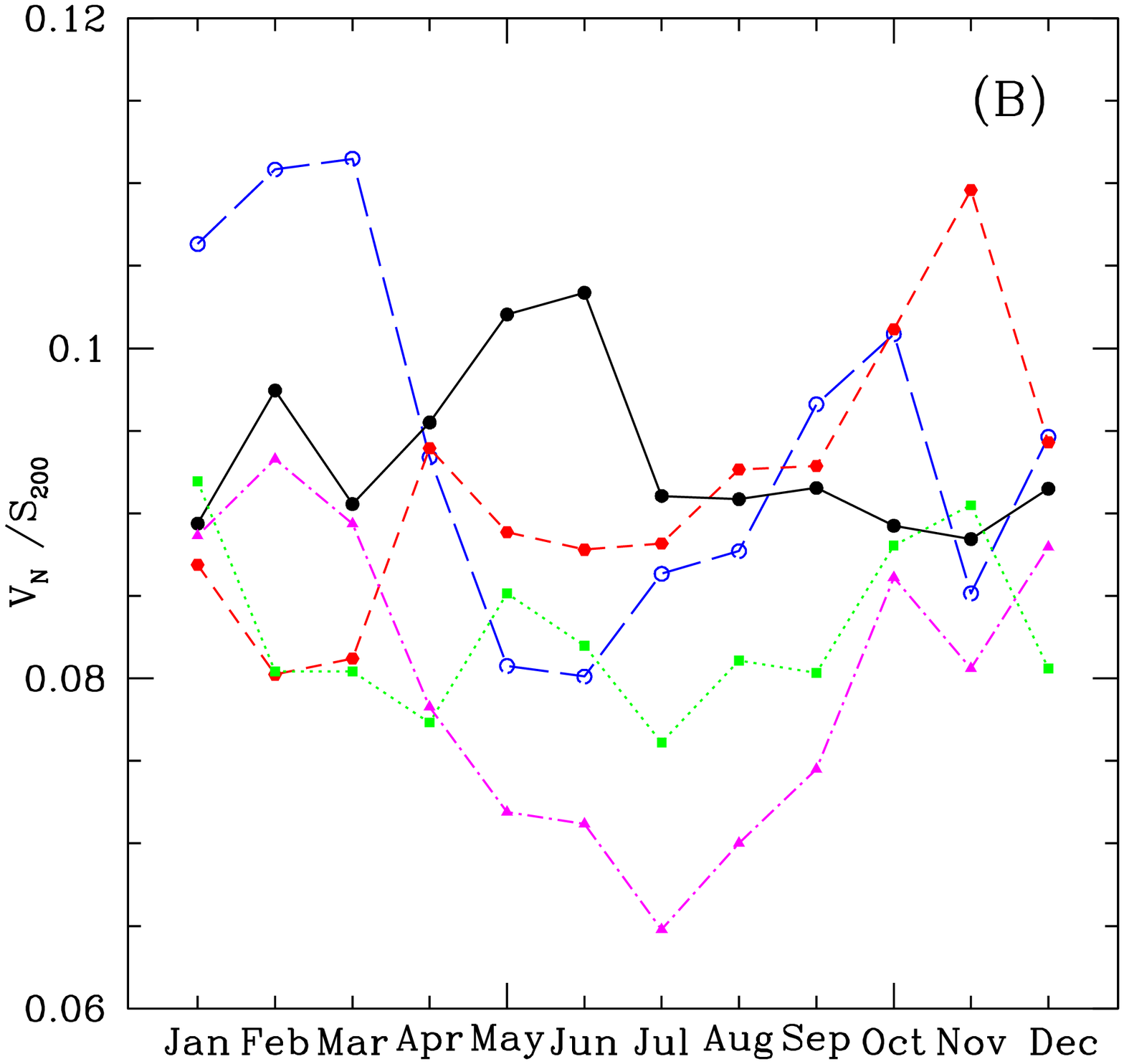}

\caption{(A) The monthly average wind velocity at 200-mbar pressure level for the
period 1980-2002 for the five selected observing sites.  (B) Normalized mean to standard
deviation ratio of wind velocity at 200-mbar pressure level for the period
1980-2002 for the selected observing sites.  In panels, lines and symbols indicate
sites as in figure \ref{pearson}.}

\label{statistic1}
\end{figure}

We have derived the monthly average, $V_{200}$, for the period 1980--2002 for the
five selected sites, and the resulting monthly statistics are shown in 
figure \ref{statistic1}A.  There is a seasonal trend in the 
high altitude wind behaviour.  In the northern hemisphere, 
the highest $V_{200}$ occurs during spring and the
lowest in summer.  La Silla and Paranal in the southern  hemisphere show
high V$_{200}$ values during the southern winter and spring.  

In the northern hemisphere, San
Pedro M\'artir shows the lowest $V_{200}$ during the summer, while the lowest
$V_{200}$ in winter corresponds to La Palma.  In autumn, Mauna Kea and La Palma
show similar values of $V_{200}$.  The differences in $V_{200}$ are small for the
 three northern
observatories  during the spring.  We define $V _{N}$ as the ratio of the
monthly average wind velocity at 200 mbar to the V$_{200}$ mean over the full 
period (Table
\ref{global})  ($V_{N}=V_{\rm 200monthly}/V_{\rm 200globalmean}$).  This ratio
has the same seasonal behaviour as $V_{200}$ at each observatory but magnitudes
are related to the same reference level, the V$_{N}$ mean, which is always
equal to unity.  If $S_{200}$ is the monthly standard deviation of $V_{200}$, the
$V_{N}$ to $S_{200}$ ratio gives a view of the stability of high altitude winds,
as shown in 
Figure \ref{statistic1}B. 
  The $V_{N}/S_{200}$ ratio at any site are directly
comparable, with higher $V_{N}/S_{200}$ values indicating more stable $V_{200}$
regimes.  According to the defined stability parameter, $V_{N}/S_{200}$, San
Pedro M\'artir presents the least stable high altitude winds of the three
northern sites.  La Palma shows a similar
stability parameter throughout the year, with the more stable $V_{200}$ during May
and June.  Mauna Kea presents a seasonal behaviour of the stability parameter,
with  highest values in February and March and lowest during May and June.
Although the high altitude winds at Mauna Kea are more stable than at La Palma
from January to March, $V_{200}$ is higher in Mauna Kea during these months.  Only
September and October shows a better behaviour at Mauna Kea than at La Palma, with
similar $V_{200}$ values but high stability at Mauna Kea.  

In the south
hemisphere, Paranal has more stable high altitude winds than La Silla except in
January.  At Paranal, November is the most stable month in terms
of $V_{200}$.  Table
\ref{global} lists the $V_{200}$ mean, median, standard deviation, and
root mean square values for the period 1980--2002 for the five 
observatories studied.  

In terms of $V_{200}$, La Palma is the best site of the five 
studied, with a mean value of 22.13 m s$^{-1}$, the smallest standard deviation (Table
\ref{global}) and showing the most stable high altitude winds.  In contrast,
La Silla shows the highest average value, with a mean $V_{200}$ of 33.35 m s$^{-1}$,
although the amplitude in Table \ref{global} (the difference between minimum and
maximum) at this site is only 12.46 m s$^{-1}$, the lowest of the five observatories
studied.  The statistical $V_{200}$ amplitudes for Paranal and Mauna Kea are quite
similar, while the value for San Pedro M\'artir is approximately twice the
 $V_{200}$ amplitude at La Silla or La Palma.  According to the
relationship connecting $V_{0}$ and $V_{200}$ \citep{sara2001}, large $V_{200}$
amplitudes indicate widely varying turbulence conditions over the course of the year.

\begin{table*}
\centering
 \begin{minipage}{180mm}
  \caption{Statistical results of 200 mbar wind speed (m s$^{-1}$) by using the data from NCEP/NCAR Reanalysis database for the period 1980--2002 at different astronomical sites. Amplitude is the difference of maximum and minimum statistical winds.}
\begin{tabular}{l|ccccc}
{\bf Site} & {\bf Amplitude (m s$^{-1}$)} & {\bf Mean (m s$^{-1}$)} & 
{\bf Median (m s$^{-1}$)} & {\bf Standard deviation (m s$^{-1}$)} & 
{\bf RMS (m s$^{-1}$)} \\ \hline
La Palma &  13.69 & 22.13  & 20.79 & 11.67 & 0.06 \\ 
La Silla &  12.46 & 33.35  & 32.77 & 12.94 & 0.07 \\ 
Mauna Kea & 18.00 & 24.33  & 22.81 & 12.30 & 0.07 \\
Paranal &   18.47 & 30.05  & 28.63 & 13.01 & 0.07 \\
San Pedro & 26.49 & 26.55  & 24.57 & 15.39 & 0.08 \\ \hline
\end{tabular}
\label{global}
\end{minipage}
\end{table*}

\begin{figure}
\includegraphics[scale=0.9]{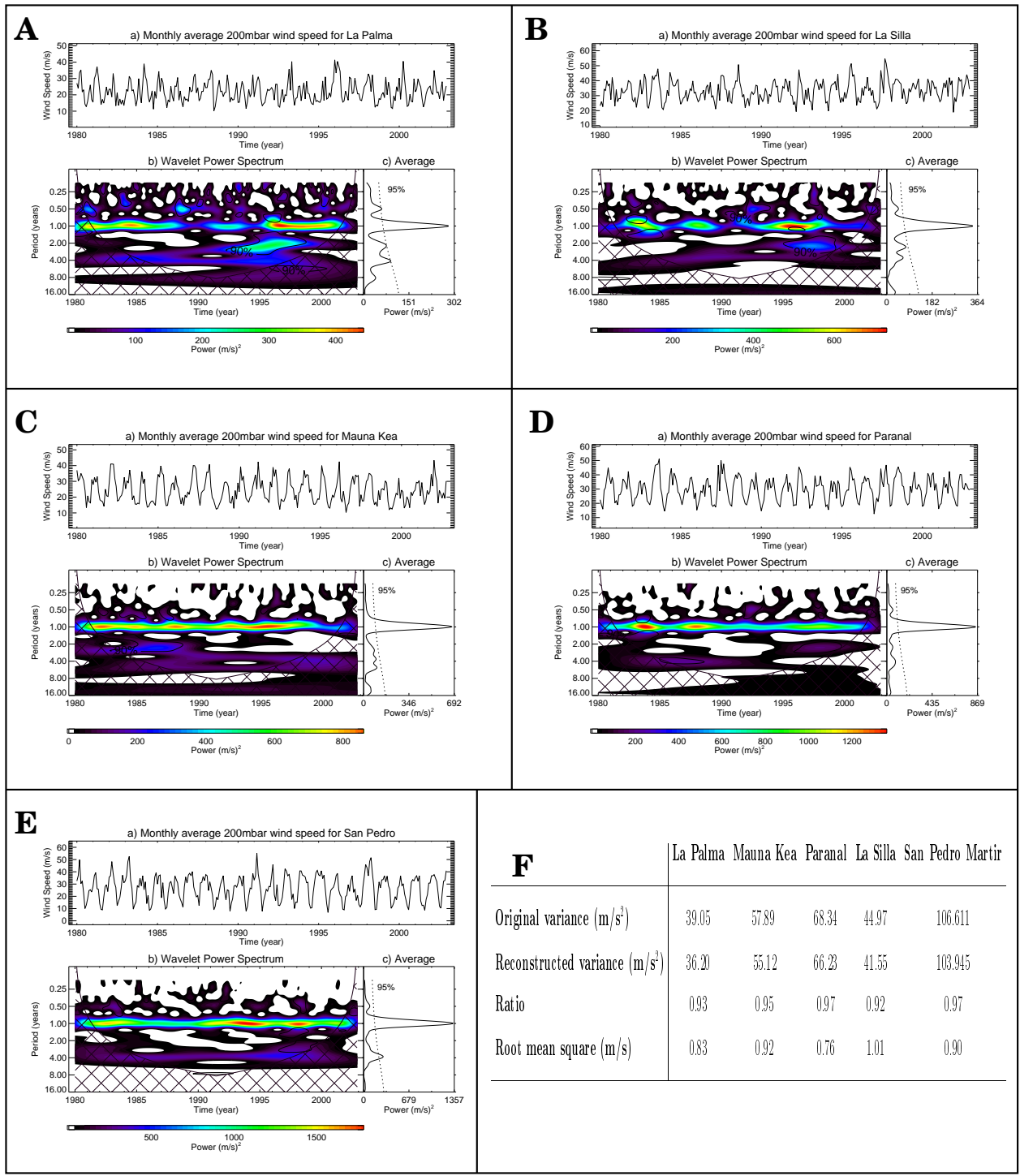}

\caption{Wavelet analysis results for the mean monthly 200 mbar wind velocity at:
 (A) ORM (La Palma), Spain; (B) La Silla, Chile; (C) Mauna Kea, Hawaii, USA; (D)
 Paranal, Chile; and (E) San Pedro Martir, Mexico. In these plots  a) is the time
 series used for the wavelet analysis; b)  is the local wavelet power spectrum of the
 time-series obtained from the Morlet wavelet.  The colour scale at the bottom
 indicates the power spectrum intensity of the full range; plot c) is the average
 power spectrum.  The dashed line is the 95 per cent confidence level.  Panel (F)
 show the results comparing the reconstruction time series from the wavelet
 analysis and the input time series.}

\label{wavelet}
\end{figure}

In order to determine any dominant temporal variation of $V_{200}$ over the
observatories in question, we have performed a wavelet analysis (see \citet{to98} for
a practical step-by-step description of wavelet analysis) of the time series
obtained from NCEP/NCAR databases for the five selected observatories.  We used a
Morlet function as wavelet mother.  Figure \ref{wavelet} shows the time series
(plot a), the local wavelet power spectrum (plot b) and the time average power
spectrum (plot c) corresponding to the wavelet analysis of each of the five
observatories in study.  A table of the reconstruction time series accuracy for
each site is shown in panel F of Figure \ref{wavelet}. The wavelet analysis of
$V_{200}$ for La Palma site has already been discussed by Chueca et al.\ (2004), and 
is in good agreement with the present results.

The annual periodicity of $V_{200}$ remains significant for all the five sites 
studied.  Except for Mauna Kea, a 0.5 years period appears in the global power
spectrum for the other four selected sites, although only for La Palma is this peak
 still significant in the local wavelet power.  For La Palma and La Silla, a 2.3
 years period is also significant in the local power spectrum, but they are less than
significant globally.  A four year period appears for the five observatories in
study, but only in San Pedro M\'artir does it remain significant in the global power
spectrum.

Atmospheric conditions at the Paranal site has  changed considerably from site 
testing period in the early 1990s to the present (ESO, 2001). The increase in  the 
average seeing  from 1998 onwards at Paranal has been identified with El Ni\~no/La 
Ni\~na events in South America (Beniston, Casals, \& Sarazin 2002). However, we 
have  found no similar discrepancy between high altitude winds before and after 
1998 at Paranal. The temporal evolution of the annual periodicity of $V_{200}$ 
remains quasi-constant from 1990 (Figure \ref{wavelet}D). The mean and standard 
deviation of $V_{200}$ for the period 1990--4 are 29.19 m s$^{-1}$ and 7.66 m 
s$^{-1}$, 
respectively. These values are slightly lower than those obtained for the period 
1998--2002 (30.77 m s$^{-1}$ and 8.31 m s$^{-1}$
 for the mean and standard deviation of $V_{200}$, 
respectively) but they are still in agreement. Therefore, the influence of events 
affecting the atmospheric conditions at Paranal site at low altitude has no 
counterpart to  winds at high altitude.

\subsection{High and low altitude wind connection}

Although only wind speed at the 200 mbar pressure level has been proposed as a
parameter for site characterization, it seems reasonable to assume 
that statistical wind
profiles could also be important in determining stability and the connection of high 
altitude
winds to turbulence at  ground level.  As a first approach,  we
explore the connection of high to lower altitude atmospheric winds by using 
the  NCEP/NCAR database.  We consider the pressure levels in the
troposphere over the observing sites available from the NCEP/NCAR database.  These
pressure levels are 700, 600, 500, 400, 300, 250, 200, 150 and 100 millibars for
La Palma, La Silla, Paranal and San Pedro M\'artir.  For Mauna Kea, the 700 mbar
pressure level is always below the  altitude of the observatory.  
The connection between the
boundary layer and high altitude winds is the subject of a forthcoming 
study (Varela et al., in preparation).

\begin{figure}
\centering
\includegraphics[scale=0.75]{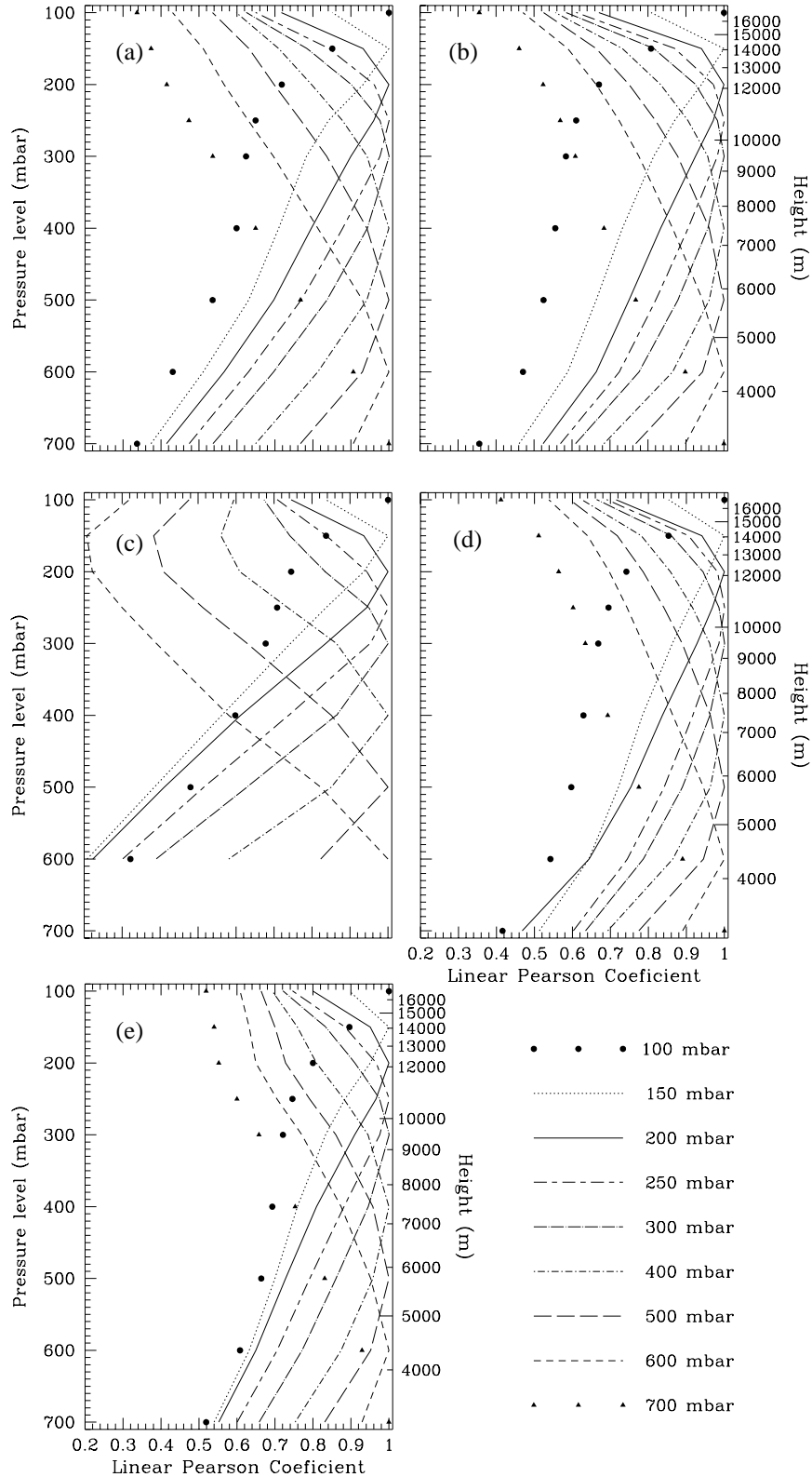}
\vspace{0.1cm}
 \caption{ Linear Pearson correlation coefficient obtained comparing the wind data 
 at each pressure level to the wind measurements at any other level at: 
 (a) La Palma, (b) La Silla, (c) Mauna Kea, (d) Paranal and (e) San Pedro Martir. 
 The comparison level is indicated by a different line type or symbol according to 
 the bottom-right table. The value equal to 1  also indicates the comparison 
 pressure level.}
\label{pearson_levels}
\end{figure}
 
In order to reveal the relationship of tropospheric winds at different altitudes,
we have compared the 6-hourly and daily wind measurements at each pressure level
with  simultaneous data at any other levels.  Figure \ref{pearson_levels}
shows the results of this comparison. The lowest levels, 700 mbar (600 mbar
 for Mauna Kea), are most probably affected by the topography of the sites,
 which may be breaking the linear relationship. At Mauna Kea, the linear 
 relationship
of 150 and 200 mbar to 500 mbar is lower than 0.5, suggesting a weak connection
between high and low altitude winds.  At the other sites, there
is a clear connection of high to low altitude winds, with Pearson coefficients
greater than 0.5 for any level.  These coefficients indicate a good linear
relationship between the wind measurements at different pressure levels.  Only the
comparison of highest, 100 mbar, to the lowest, 700 mbar (600 mbar for Mauna Kea)
level gives Pearson coefficients less than 0.5.  Winds at the lowest 
levels could be affected by the surrounding site topography, perhaps disturbing the
linear relationship.  In fact, the 700 mbar pressure level corresponds to
approximately 3120 metres, less than one kilometre above La Palma, La Silla,
Paranal and San Pedro M\'artir.  For Mauna Kea, the mean altitude of the
lowest pressure level, 600 mbar, is around 4350 metres, just 250 metres above the
summit.  At the other pressure extreme, 100 mbars corresponds on average to 16\,750
metres, approximately at the average altitude of the
tropopause layer (Seidel et al.\  2001; Garc\'{\i}a-Lorenzo, Fuensalida, \& Eff-Darwich 2004).  Therefore, the breaking of the linear
relationship of 100 mbar pressure level to lower altitudes could
be related to the fact that 100 mbar could be below the tropopause.  In
other words, winds measurements at 100 mbars could correspond to tropospheric,
tropopausal or mesospheric winds, depending on the season.

\begin{table*}
\centering
 \begin{minipage}{180mm}
  \caption{Mean linear relationship of wind measurements at different pressure levels. Mean corresponds to the average of Pearson coefficients obtained comparing the referred pressure level to any other.$\sigma_{m}$ indicates the standard deviation of these coefficients. Cov. is the average covariance of individual covariance values obtained comparing each level with the others, while $\sigma_{cov}$ is the standard deviation of these covariances.    }
\begin{tabular}{c|c|c|c|c|c}
{\bf Pressure } & {\bf          La Palma} & {\bf La Silla} & {\bf Mauna Kea} & {\bf Paranal} & {\bf San Pedro} \\ 
 {\bf (mbar)}         & Mean \ $\sigma_{m}$ \ Cov. \ $\sigma_{cov}$ &  Mean \ $\sigma_{m}$ \ Cov. \ $\sigma_{cov}$ & Mean \ $\sigma_{m}$ \ Cov. \ $\sigma_{cov}$ & Mean \ $\sigma_{m}$ \ Cov. \ $\sigma_{cov}$ & Mean \ $\sigma_{m}$ \ Cov. \ $\sigma_{cov}$  \\ \hline

     100 & 0.59 \  0.16 \  42.2 \  21.0 &  0.57 \  0.13 \   45.4 \  21.0 &  0.62 \   0.17 \   49.7 \  27.4  &  0.64 \  0.13 \  52.9 \  27.0 & 0.70 \  0.11  \  77.7 \   36.3 \\
   150 & 0.70  \  0.18 \   64.4 \   33.0 &  0.73 \   0.15 \    79.5 \   39.5 &  0.64 \   0.26 \  69.6 \  45.0  &  0.78 \  0.14 \  86.2 \  44.9 & 0.77 \  0.14 \  115. \   57.8 \\
   200 & 0.74  \  0.18 \   76.2 \   38.0 &  0.78 \   0.15 \    97.7 \   47.4 &  0.67 \   0.27 \  76.3 \  48.4  &  0.78 \  0.17 \  98.1 \  49.7 & 0.79 \  0.14 \  132. \   65.3 \\
   250 & 0.77 \   0.17 \   76.8 \   36.9 &  0.80 \   0.15 \   100. \   47.5 &  0.71 \   0.23 \  74.3 \  43.4  &  0.83 \  0.14 \  97.0 \  48.3 & 0.81 \  0.13 \  131. \   63.1 \\
   300 & 0.78 \   0.15 \   71.7 \   32.7 &  0.81 \   0.15 \    95.3 \   44.1 &  0.72 \   0.18 \  65.3 \  34.0  &  0.84 \  0.13 \  89.7 \  43.9 & 0.83 \  0.11 \  121. \   55.6 \\
   400 & 0.79 \   0.12 \   58.2 \   22.7 &  0.81 \   0.14 \    78.7 \   33.6 &  0.68 \   0.13 \  43.6 \  16.1  &  0.83 \  0.12 \  72.3 \  32.7 & 0.83 \  0.09 \   97.8 \   38.8 \\
   500 & 0.76 \   0.14 \   45.5 \   14.4 &  0.78 \   0.14 \    61.3 \   23.1 &  0.58 \   0.19 \  26.6 \   5.80  &  0.81 \  0.12 \  55.8 \  22.2 & 0.81 \  0.11 \   76.4 \   25.6 \\ 
   600 & 0.68 \   0.18 \   33.1 \   8.34  &  0.74 \   0.16 \    45.4 \   14.3 &  0.40 \   0.22 \  13.8 \   3.10  &  0.76 \  0.13 \  38.2 \  12.4 & 0.76 \  0.13 \   54.8 \   14.5 \\
   700 & 0.55 \   0.20 \   22.1 \   4.84  &  0.60 \   0.17 \    28.1 \    7.32 &                                 &  0.63 \  0.15 \  20.1 \   4.70 & 0.67 \  0.15 \   34.9 \    6.55 \\ \hline

\end{tabular}
\label{mean_pearson_levels}
\end{minipage}
\end{table*}

Table \ref{mean_pearson_levels} shows the mean  values of
winds at different levels, derived by averaging the individual Pearson coefficients
obtained from comparing a particular level to any other (figure \ref{pearson_levels})
and the standard deviation of these coefficients ($\sigma_{m}$).  We have not
considered the comparison of any level to itself (Pearson coefficient of 1 in
Figure \ref{pearson_levels}).  Table \ref{mean_pearson_levels} also includes the
mean covariance (Cov.)  and its standard deviation ($\sigma_{\rm cov}$) as an
indicator of the strongest of the linear relationships.  Although the largest mean
correlation coefficients corresponds to 400 mbar above La Palma, La Silla and San
Pedro M\'artir, and 300 mbar for Mauna Kea and Paranal, the maximum mean covariance
corresponds to a level of around 200 mbar.

Combining Figure \ref{pearson_levels} and results of table
\ref{mean_pearson_levels}, we may conclude that high altitude winds show good
relationships to low altitude winds.  The highest correlation corresponds to
Paranal and San Pedro M\'artir.  La Palma and La Silla show a similar behaviour in
this regard, while Mauna Kea shows the worse correlation between high to low winds.
On the basis of  these results, we  deduce that the relationship V$_{0}\propto
V_{200}$ (Sarazin \& Tokovin 2001) should be worse for Mauna Kea than for any of
the other sites considered in this paper.  The topographical influence seems to be
slightly greater at the sites on islands than at the continental sites.

\subsection{Wind profile statistics and seasonal behaviour}

\begin{figure}
\includegraphics[scale=0.75]{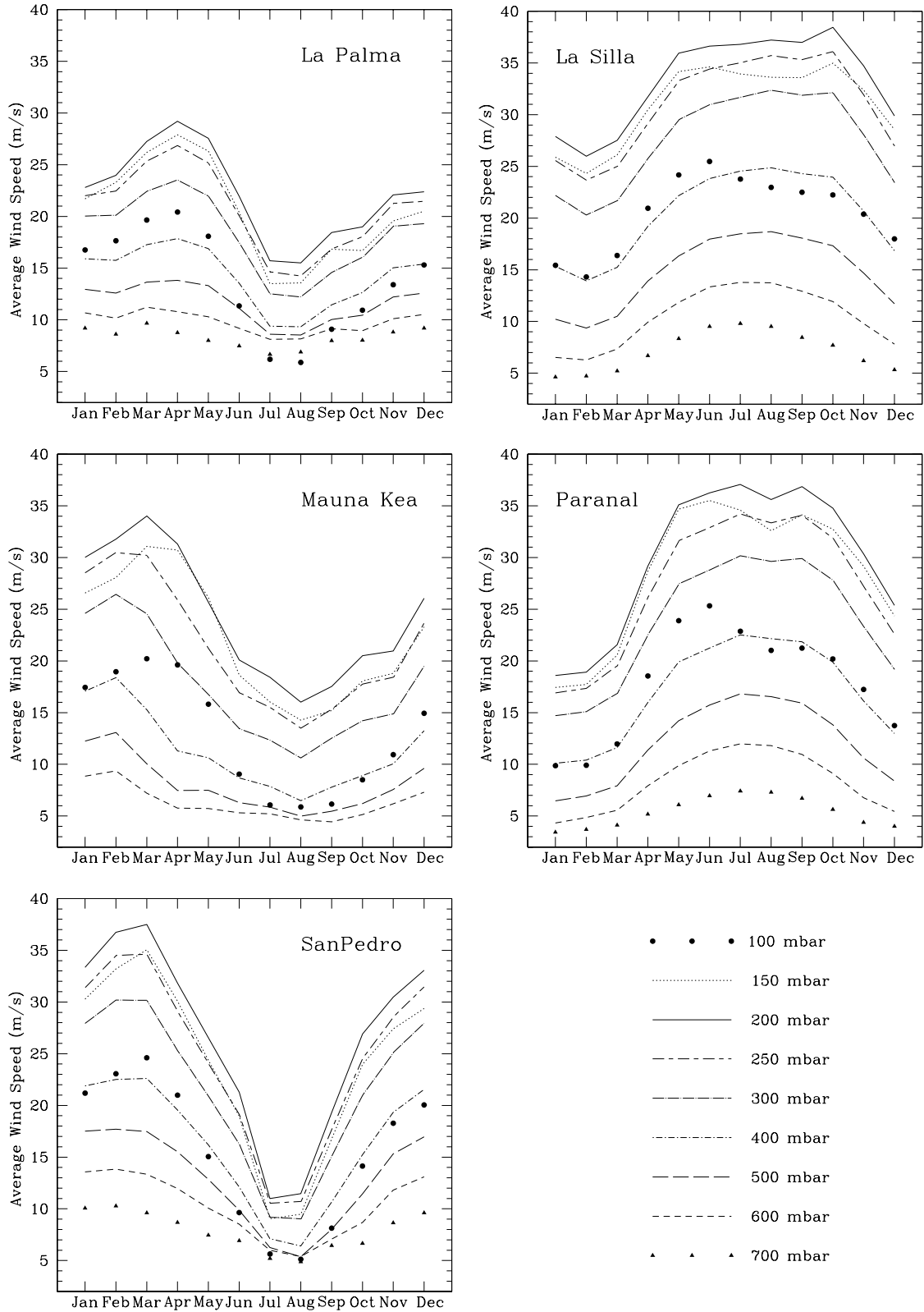}
 \caption{The monthly average wind velocity for the period 1980--2002 at the 
 pressure levels indicated in the bottom-right table for the selected sites. }
\label{statistic_height}
\end{figure}

Figure \ref{statistic_height} shows the monthly average wind for the period
1980--2002 at different pressure levels for the five  observatories.  Wind
statistics at the different pressure levels follow a similar seasonal behaviour
to that of $V_{200}$ (Section 3.1),  $V_{200}$ being the highest
 winds at any month and
site.  Only Mauna Kea shows slightly stronger mean winds at 150 mbar than
$V_{200}$ during May.  The monthly average winds at 100 mbar show the largest
differences from the statistics of all the others at the 
 pressure levels considered, perhaps
 because the 100 mbar level can be above or below the tropopause.  The
 observatories studied seem to be statistically stable in the sense that they
show the worst and best wind measurements
for the same epoch at any of the  pressure
levels considered.  
Only Mauna Kea shows the strongest winds in levels at different months:
while the statistical $V_{200}$ peak corresponds to March, the strongest
 winds for 150 and 250 mbar occur in April and February, respectively.
The monthly average wind behaviour shown in Figure \ref{statistic_height} for
Mauna Kea confirms the results of the previous section, showing that the
connection between high and low altitude winds is not clear for the Mauna Kea site.

Table \ref{global_levels} shows the annual statistical results of the wind
profiles for the five  observatories.  Vertical wind  gradients are larger
in winter than in summer, indicating   stabler and hence less turbulent 
atmospheres
in summer all the  observatories studied.

\begin{table*}
\centering
 \begin{minipage}{130mm}
  \caption{Statistical results of wind speed profile (m s$^{-1}$) by 
  using the data from NCEP/NCAR Reanalysis database for the period 1980--2002 
  at different astronomical sites. Amplitude is the difference of maximum and 
  minimum statistical winds at each pressure level.}
\begin{tabular}{l|cccccc}
{\bf Site} & {\bf Level } & {\bf Amplitude}    & {\bf Mean }        & {\bf Median}       & {\bf Standard deviation} & {\bf RMS} \\ 
           & {\bf (mbar)} & {\bf (m s$^{-1}$)} & {\bf (m s$^{-1}$)} & {\bf (m s$^{-1}$)} & {\bf (m s$^{-1}$)}            & {\bf (ms$^{-1}$)} \\ \hline
         &  100   &  14.55   & 13.69 & 12.69 &  7.72 &  0.04  \\
         &  150   &  14.39   & 20.51 & 19.43 & 10.24 &  0.06  \\
         &  250   &  12.61   & 20.66 & 19.04 & 11.49 &  0.06  \\
{\bf La Palma} &  300   &  11.30   & 18.23 & 16.56 & 10.49 &  0.06  \\
         &  400   &   8.53   & 14.18 & 12.71 &  8.44 &  0.05  \\
         &  500   &   5.29   & 11.64 & 10.42 &  6.86 &  0.04  \\
         &  600   &   3.09   &  9.78 &  8.79 &  5.63 &  0.03  \\
         &  700   &   3.02   &  8.27 &  7.52 &  4.64 &  0.03  \\ \hline
         &  100   &  11.17   & 20.58 & 20.37 &  7.73 &  0.04  \\
         &  150   &  10.68   & 31.09 & 30.74 & 10.93 &  0.06  \\
         &  250   &  12.42   & 31.04 & 30.23 & 13.05 &  0.07  \\
{\bf La Silla} &  300   &  12.06   & 27.52 & 26.70 & 12.20 &  0.07  \\
         &  400   &  10.92   & 20.45 & 19.48 & 10.01 &  0.05  \\
         &  500   &   9.32   & 14.80 & 13.73 &  7.96 &  0.04  \\
         &  600   &   7.50   & 10.46 &  9.35 &  6.27 &  0.03  \\
         &  700   &   5.18   &  7.20 &  6.16 &  4.68 &  0.03  \\ \hline
         &  100   &  14.14   & 12.76 & 10.85 &  8.35 &  0.05  \\
         &  150   &  16.81   & 22.22 & 20.54 & 11.53 &  0.06  \\
         &  250   &  16.98   & 21.39 & 19.77 & 11.35 &  0.06  \\
{\bf Mauna Kea} &  300   & 15.83   & 17.41 & 15.73 &  9.83 &  0.05  \\
         &  400   &  11.88   & 11.27 &  9.75 &  7.12 &  0.04  \\
         &  500   &   8.09   &  8.00 &  6.85 &  5.25 &  0.03  \\
         &  600   &   4.93   &  6.24 &  5.46 &  3.99 &  0.02  \\ \hline
         &  100   &  15.45   & 18.05 & 17.83 &  8.35 &  0.05  \\
         &  150   &  18.09   & 28.58 & 27.99 & 11.73 &  0.06  \\
         &  250   &  17.29   & 27.38 & 25.76 & 12.66 &  0.07  \\
{\bf Paranal}  &  300   &  15.46   & 23.85 & 22.29 & 11.54 &  0.06  \\
         &  400   &  12.41   & 17.13 & 15.70 &  9.29 &  0.05  \\
         &  500   &  10.34   & 12.12 & 10.66 &  7.30 &  0.04  \\
         &  600   &   7.67   &  8.39 &  7.20 &  5.31 &  0.03  \\
         &  700   &   3.98   &  5.43 &  4.80 &  3.31 &  0.02  \\ \hline
         &  100   &  19.49   & 15.45 & 14.27 &  9.59 &  0.05  \\
         &  150   &  26.01   & 23.94 & 22.77 & 13.54 &  0.07  \\
         &  250   &  24.07   & 24.62 & 22.24 & 14.87 &  0.08  \\
{\bf San Pedro} &  300   & 21.15   & 21.44 & 18.88 & 13.45 &  0.07  \\
         &  400   &  16.22   & 16.22 & 13.88 & 10.77 &  0.06  \\
         &  500   &  12.32   & 12.83 & 10.82 &  8.63 &  0.05  \\
         &  600   &   8.45   & 10.26 &  8.71 &  6.52 &  0.04  \\
         &  700   &   5.40   &  7.86 &  6.93 &  4.67 &  0.03  \\ \hline
\end{tabular}
\label{global_levels}
\end{minipage}
\end{table*}

\section{Discussion and Conclusion}

The connection between $V_{200}$ and the average velocity of the turbulence, $V_{0}$,
found by Sarazin \& Tokovinin (2001) at Paranal and Cerro Pach\'on has been
accepted as a parameter for tackling astronomical site suitability for adaptive 
optics.  The linear relationship between $V_{0}$ and the vertical wind 
profile, $V(h)$, is established by the definition of $V_{0}$:

\begin{center}
\begin{equation}
V_0=\frac{\int_{0}^{\infty}C_{N}^{2}(h) V(h) dh}{\int_{0}^{\infty}
C_{N}^{2}(h) dh},
\end{equation}
\end{center}
where $C_{N}^{2}(h)$ is the turbulence vertical profile.  The high
level of agreement between winds at any altitude demonstrated at  Section 3.2
therefore suggests a relationship between $V_{0}$ and $V_{200}$ at any of the 
observing sites studied. However, we have only demostrated a connection of 
high and 
low altitude winds, but not a connection of high altitude winds and the average 
velocity of the turbulence. Even if such a connection were valid 
world-wide, it could be either linear or non-linear. In the mean time we  consider a 
linear relationship at the  sites studied similar to that found at Paranal and 
Cerro Pach\'on. Our results support the idea of considering $V_{200}$ as a
parameter for ranking astronomical sites with regard to
 their suitability for adaptive optics in
spite of the lack of information on $C_{N}^{2}(h)$. However, the importance of
deriving atmospheric turbulence profiles is still essential.  $C_{N}^{2}(h)$
determines the coefficient of the linear fit connecting V$_{0}$ and V$_{200}$ (A
hereafter, $V_{0}={A}\times V_{200}$).  For Paranal and Cerro Pach\'on
 this coefficient is {\it A} = 0.4 (Sarazin \& Tokovinin 2001), but
there are  still no measurements for any other site.  According to the results of
this paper, we do not have any evidence for adopting the same coefficient at the
different sites.  Even in the hypothetical case of similar $C_{N}^{2}(h)$ at
different summits (which is less than probable), the connection between high to low
altitude winds reveals several differences at different sites, suggesting a
different {\it A} coefficient in the linear fit.  The final value of the linear
coefficient at a particular site could drastically change  the interpretation of
$V_{200}$ as a direct parameter for site characterization.  If we consider a
site with $V_{200}$ statistics similar to Paranal and {\it A} adopts larger values
than 0.4 at that particular site, we  obtain  larger $V_{0}$ values than for
Paranal.  In contrast, a smaller {\it A} value will indicate  less
atmospheric turbulence at that hypothetical site than at Paranal.

Another important point is the seasonal behaviour of $V_{200}$.  Changes in
$V_{200}$ regimes in different seasons could have an important influence on the
value of {\it A},  taking into account that we have  presented here only the
statistics of wind modules, and that we have not considered  the wind direction,
which could point to changing wind regimes with season.  We could obtain
different {\it A} values for different months; that is, {\it A} could also present
 seasonal behaviour.  To have the best parameterization of the average wind 
turbulence, $V_{0}$, in term of $V_{200}$ seems to be mandatory in any study of the
seasonal behaviour of the linear coefficient {\it A}.  Unfortunately, our
information about {\it A} is very poor, with measurements only at Paranal and
Cerro Pach\'on (Sarazin \& Tokovinin 2001).  If {\it A} were a constant
coefficient and had a similar value to that found for Paranal and Cerro
Pach\'on,  $V_{0}$ would present a similar seasonal behaviour to that of $V_{200}$.
In this hypothetical case, the statistical results obtained here for
$V_{200}$ would indicate that La Palma is at the top of the suitability ranking
list for
adaptive optics, while La Silla is at the bottom.

Atmospheric stability, in the sense of similar behaviour and good
correlation of winds at different altitudes, is also an important point in
establishing $V_{200}$ as a direct parameter for site characterization and adaptive
optics suitability.  Our results show that Paranal and San Pedro M\'artir present
statistically more stable atmospheres than the other three  sites considered.
The stability results are found similar for La Silla  and La Palma.  
Mauna Kea shows the worst results with regard to
high to low altitude winds.  Taking into account the  behaviour of $V_{200}$
 and its
connection to low altitude winds, Mauna Kea should present unstable average wind
 turbulence compared to the other sites.  However, the stability of winds at different altitudes is
not a guarantee of stable turbulence behaviour.  If $C_{N}^{2}(h)$ is stable
enough at a site with unstable winds behaviour in altitude, it should be a similar
or better site for adaptive optics than a  site with stable winds and unstable
$C_{N}^{2}(h)$ behaviour.  A more stable atmosphere, in the sense established in
this paper, suggest a better relationship between $V_{0}$ and $V_{200}$ and better
accuracy of the determination of the {\it A} coefficient  than sites with unstable
atmospheres.

Although a correlation between seeing and winds at the 200 mbar pressure 
level has not been found at any site (Carrasco \& Sarazin 2003; Sarazin \& 
Tokovinin 2001) the mistaken idea of identifying high altitude winds 
with turbulence has become increasingly popular among 
 the astronomical community. If there is indeed a relationship between  $V_{0}$ and 
$V_{200}$  at the different sites, this does not mean that the level of 
turbulence depends on $V_{200}$ but only that the coherence time will vary with 
$V_{200}$ behaviour. We wish to alert the astronomical community concerning this false 
interpretation of $V_{200}$ as a parameter related to turbulence.

\section{Summary and Conclusions}

We have analysed the wind vertical profiles for the period 1980--2002 using the 
NCEP/NCAR Reanalysis archive at  five astronomical sites. We have 
verified the consistency of the statistical results derived from these databases 
and balloon measurements. Our main results and conclusions may be summarized as 
follows:

\begin{itemize}
\item[1.] The excellent statistical correlation between NCEP/NCAR Reanalysis data 
and balloon measurements reveal the NCEP/NCAR Reanalysis archive to be a useful source 
of meteorological data for site characterization.

\item[2.] In terms of high altitude wind speed ($V_{200}$) statistics, 
La Palma is the 
best  of the five site studied, with a mean value of 22.13 m s$^{-1}$, the lowest 
standard deviation and the most stable V$_{200}$ behaviour. La Silla is at the 
botton of the ranking with a mean $V_{200}$ of 33.35 m s$^{-1}$, although the  
seasonal $V_{200}$ amplitude is the lowest of the five cases, with a statistical 
value of 
12.46 m s$^{-1}$.

\item[3.] We have found a clear annual periodicity of $V_{200}$ at the five  
observatories.

\item[4.] We have found a good level of correlation between high and low altitude 
winds, supporting a linear relationship between $V_{200}$ and the average velocity 
of the turbulence at any of the sites. However, the proportional coefficient of 
this relationship may differ from site to site as a result of the clear seasonal 
behaviour of winds and the large differences of wind vertical profiles behaviour 
at the  sites.

\end{itemize}

The results in this paper does not confirm a relationship of V$_{0}$ and 
V$_{200}$ at the studied sites, but only a connection of high and low altitude 
winds. V$_{0}$ strongly depends on turbulence structure, and therefore, its 
connection to high altitude winds could be only stablish with a proper turbulence 
characterisation. We alert the astronomical community on the wrong idea of 
identify high altitude winds and turbulence at the ground level that it is being 
more and more popular in the recent meetings. If a V$_{0}$ and V$_{200}$ 
relationship were valid elsewhere, we may connect coherence time and high altitude 
winds, but never image quality.

\section*{Acknowledgments}

The authors thank T. Mahoney for his useful discussions and assistance.
 This work has made use of the NCEP Reanalysis data provided by the NOAA-CIRES
 Climate Diagnostics Center, Boulder, Colorado, USA, from their Web site at
 http://www.cdc.noaa.gov/. Wavelet software was provided by C. Torrence and
 G. Compo, and is available at URL: http://paos.colorado.edu/research/wavelets/.
 This work has been partially funded by the Spanish Ministerio de Ciencia y 
 Tecnología (AYA2003-07728).

\end{document}